\documentclass{iucr}              
\usepackage{epsfig,color}
\usepackage{graphicx}
\usepackage{amsmath,amssymb}

     \journalcode{}              

\begin{document}                  



\title{About the coherent fraction of synchrotron emission}


\cauthor{Manuel}{Sanchez del Rio}{srio@esrf.eu} 

\aff{ESRF, 71 Avenue des Martyrs, 38043 Grenoble \country{France}}









\maketitle                        

\begin{synopsis}
An accurate definition of the coherent fraction is proposed, with a recommendations of the best way to approximately calculate it using different equations from literature.
\end{synopsis}

\begin{abstract}
Coherent fraction is a concept widely used in the synchrotron radiation community. Here I discuss the meaning of coherent fraction, an accurate way to define it, and the different formulations found in literature to calculate it. After several comparisons using values from the present and future ESRF lattices, an approximated way to quickly compute it is proposed. 
\end{abstract}


\section{Introduction}

Most reports of new storage ring light sources and upgraded sources include plots of the 'coherent fraction' or 'coherence fraction' to make evidence an important gain in coherence. Coherent fraction would account for the number of transverse spatial 'coherent photons' over the total number of photons emitted by the source at a given photon energy. There is however a lack of rigor and information not only on how to obtain it, but also in its definition. Different formulas are found in literature which are not always compatible and cause confusion in the synchrotron community. See, for example, different equations for photon emission in \cite{walker}. I try here to describe in detail the coherent fraction, comment the basic references in the field, discuss about the confusion created by the coexistence of different approximations, and give some recommendations on which equations are preferred to be used. 

I introduce first the case of Gaussian beams, and then the beams produced by undulators in a storage rings. The discussion regards always the use of the phase space for describing the electron beam in the accelerator and the photon beam in beamlines. 

\section{About 'divergence' in optical wavefronts}

The phase space method is well adapted to describe statistically the electrons circulating in a storage ring. Electrons follow well defined particles where  the position and the moment (or direction) are both perfectly defined (in Classical Mechanics) and can be calculated at any point of the storage ring. The Classical Electromagnetism permits to calculate the electric field emitted by one electron (or many electrons that follow exactly the same trajectory, called filament beam) in a given position. The mapping of the electric field in a plane perpendicular to the propagation direction (i.e., at a coordinate $z$ along the beamline) permits to build a wavefront which is, by definition, monochromatic and fully coherent. One can build the intensity map (square modulus of the electric field) at that position. There is in principle no 'direction' or divergence associated to a wavefront. The intensity distribution of the wavefront changes importantly  when it propagates from $z$ to $z+\Delta z$ with $\Delta z$ 'small' (near field). This is usually obtained calculating the Fresnel-Kirchhoff integral. For 'large' $\Delta z$, we enter in the so called Fraunhofer regime (far field), and the shape of the intensity does not change, only expands with distance. Therefore, knowing the intensity plot $I(x_1,y_1)$ at $z=z_1$, one can calculate the values at $z_2=z_1+\Delta z$ by multiplying the horizontal (x) and vertical (y) coordinates by $\Delta z/z_1$. It is now when the concept of divergence appears, because the intensity is invariant when expressed as a function of 'divergence' axes $x/z$ and $y/z$ (always supposing $z \gg x,y$). Fraunhofer theory states that the electric field at $z_2=z_1+\Delta z$ is proportional to the Fourier Transform of the electric field at $z_1$:

\begin{equation}
E(x_2,y_2) = \frac{-e^{-ikz}}{i \lambda z_2} e^{-\frac{i k}{2z_2}(x_2^2 + y_2^2)} \int{E(x_1,y_1)} e^{-\frac{i k}{z_2}(x_1 x_2 + y_1 y_2)} dx_1 dy_1 
\end{equation}
where $E(x,y)$ is the electric field amplitude in a given $z$ position with the conjugated variables: 

\begin{equation}
x ;  f_x = \frac{x}{\lambda z}  \sim \frac{ \theta_x }{ \lambda } 
\end{equation}
and the same for the $y$. In consequence, talking about 'divergence' of  a wavefront makes sense only if one thinks how this wavefront will propagate in the far field.

From mathematics we know that the Fourier transform of  Gaussian $g(x) = e^{-ax^2}$ is another Gaussian $G(f_x) = \sqrt{\pi/a} e^{- \pi^2 f_x^2/a}$. Expressing $a=1/(2 \sigma_{A,x}^2)$ and writing $G(f_x) = \sqrt{\pi/a} e^{- f_x^2/(2 \sigma_{A,f_x}^2)}$ one obtains that the standard deviations (sigmas) are related by 
\begin{equation}
\label{FTmath}
\sigma_{A,x} \sigma_{A,f_x}= \frac{1}{2 \pi}
\end{equation}

Fraunhofer propagation expresses the relation the electric field amplitude $E_2$ in a position $z_2$ with the electric field amplitude $E_1$ in the position $z_1$ in such a way that $E_2$ is the Fourier transform of the field distribution $E_1$ with $E_1$ expressed in real space and $E_2$ in the conjugated variable. Therefore, Eq.~\ref{FTmath} holds is the sigmas are related to the electric field amplitude. That is the reason why the sub index $A$ is used in Eq.~\ref{FTmath}. If the Gaussian sigma is chosen to describe the intensity distribution (the square of the electric field amplitude), then $\sigma_x=\sqrt{2}\sigma_{A,x}$ and $\sigma_{f_x}=\sqrt{2}\sigma_{A,f_x}$, therefore Eq.~\ref{FTmath} takes the form: 

\begin{equation}
\sigma_{x} \sigma_{f_x}= \frac{1}{4 \pi}
\end{equation}

Hereafter, unless specified, the Gaussian $\sigma$'s always correspond to intensity. 

\section{Partial coherent beams and rigorous definition of coherent fraction}

The electrons in an storage ring are statistically distributed, following (in good approximation) a Gaussian distribution in a 6-dimensional space (three spatial coordinates, two angles to define direction, and the electron energy). Many electrons therefore contribute to photon beam radiation, each one creating a wavefront. The consequence is that the overall radiation cannot be described deterministically which implies that statistical methods are needed, like for describing the electron beam. This is the origin of partial coherence of the synchrotron emission. In an ideal storage ring of zero emittance, the electrons follow a filament beam so the emission is fully coherent. As soon as  electron emittance increases, the electrons contributing to the radiation start to have different initial conditions (in the 6-dimensional space) and the overall emission cannot be describe by a single wavefront, but by statistically distributed wavefronts, described by partial coherent optics. Under some circumstances that are almost always satisfied for light emitted by storage rings (but not for XFELs) \cite{geloni2008}. They can be summarized in i) the electron bunch length is long enough, ii) radiation is monochomatized 'not too much' (like by standard monochromators), and iii) the radiation frequency is large enough. When these conditions are satisfied the radiation is 'wide sense stationary' \cite{mandel_wolf}. It is in this case all the coherent properties of the radiation can be described using the 'cross spectral density' (CSD), also called 'mutual intensity' that is a complex function that measures the correlation of the electric field in two different spatial points at a given radiation frequency. It can be mapped in a $(x,y)$ plane at a third coordinate $z$. At that plane, the CSD depends on 5 variables: 

\begin{equation}
W(x_1,y_1,x_2,y_2,\omega) = <E^*(x_1,y_1,\omega) E(x_2,y_2,\omega)>
\label{CSD}
\end{equation}

From the cross spectral density one can calculate the 'spectral density' (usually simply called 'intensity') $I(x,y,\omega)=W(x,y,x,y,\omega)$, and also the complex degree of (transverse) coherence: 

\begin{equation}
\mu(x_1,y_1,x_2,y_2,\omega) = \frac{W(x_1,y_1,x_2,y_2,\omega)}{\sqrt{I(x_1,y_1,\omega)}\sqrt{I(x_2,y_2,\omega)}}
\label{DTC}
\end{equation}
The modulus of the complex degree of coherence is one for a completely coherent beam and zero for an incoherent beam. 

An important result in partial coherence optics permits to describe the radiation as an infinite sum of independent coherent modes (in the sense of orthonormality) :

\begin{equation}
W(x_1,y_1,x_2,y_2,\omega) = \sum_{n=0}^{\infty} \lambda_n(\omega) \Phi^*(x_1,y_1,\omega) \Phi(x_2,y_2,\omega) 
\label{CMD}
\end{equation}
where $\lambda_n$ (eigenvalues) are the intensity weights and the $\Phi$ are the coherent modes (eigenfunctions). 
Some important characteristics of this coherent mode decomposition are: i) the modes are orthonormal (in the integral sense), ii) the modes maximize the spectral density, the first mode is more intense than the second, and so on, meaning that the truncated expansion is optimal, and iii) there is complete coherence if and only if there is only a single mode. If one can truncate the infinite series to a limited number of modes $N$ which contain a good percentage of the spectral density, the numerical storage of the $N$ modes that depend on two spatial variables is usually more economic than the storage of the cross spectral density $W$ that depends on four variables. 
The eigenvalues $\lambda_n$ are a measure of the intensity. One can define the occupation $\eta$ of the i-th mode as the normalized intensity: 

\begin{equation}
\eta_i(\omega) = \frac{\lambda_i(\omega)}{\sum_{n=0}^{\infty} \lambda_n(\omega)}
\end{equation}

From these arguments, it is now natural to rigorously define Coherent Fraction ($CF$) as the occupation of the first coherent mode: $CF=\eta_0$.

\section{A {\it scholar} case: the Gaussian Shell-model}

The Gaussian Shell-model has in principle nothing to do with synchrotron radiation, but it is interesting to look at it as it has been proposed as a model for synchrotron emission \cite{coisson1997,vartanyants2010}, and it is at the origin of the equations used in literature for obtaining the CF. The Gaussian Shell-model is a case of partial coherent beams very well know in Optics, because it applied quite well to lasers and it can be described analytically. It is often used as a reference model for partial coherent beams. It is a 1-dimensional model based on the hypothesis that both the spectral density (intensity distribution) and the degree of transverse coherence between two points follow Gaussian distributions, described by standard deviations $\sigma_I$ and $\sigma_{\mu}$, respectively. 

The resulting coherent modes \cite{Starikov82,mandel_wolf} are described by the sometimes called Hermite-Gaussian functions:
\begin{equation}
\Phi_n(x) = \frac{1}{\sqrt{2^n n!}} \left( \frac{2c}{\pi} \right)^{1/4} \frac{1}{\sqrt{2^n n!}} H_n(x\sqrt{2c})e^{-cx^2}
\label{GSeigenvalues}
\end{equation}
where 
\begin{equation}
c = \frac{1}{2 \sigma_I} \left[ \frac{1}{(2 \sigma_I)^2} + \frac{1}{\sigma_{\mu}^2} \right]^{1/2}
\end{equation}
and the $H_n$ are the physicist's Hermite polynomials of order $n$. 
The first mode is just a Gaussian ($\Phi_0 \propto e^{-cx^2}$), which propagates into the far-field as another Gaussian $ \propto e^{-f_x^2/4c}$ .
Similarly to a Gaussian that Fourier-transforms to a Gaussian, the coherent modes (Gaussian-Hermite modes) are also invariant by Fourier transform. Because Fraunhofer propagation is expressed as Fourier transform, the Gaussian-Hermite coherent modes do not change in shape, but only expand as they are propagated farther and farther. It is in general not possible to arrive to a similar expression for other beams that do not follow the Gaussian Shell model. 

It can be shown that the first coherent mode of the Gaussian Shell-model (see e.g., \cite{Siegman_1990}): 

\begin{equation}
\sigma_x \sigma_{\theta} = \frac{\lambda}{4 \pi}
\label{sigmasigmaprime}
\end{equation}
with $\lambda$ the photon wavelength, $\sigma_x$ is the Gaussian sigma of the spatial distribution of the intensity and $\sigma_{\theta}$ is the Gaussian sigma of the angular distribution of the intensity, certainly at the far field. 

The coherent fraction, that is the occupation of the first coherent mode, can be obtained for the Gaussian Shell-model from the distribution of the eigenvalues \cite{Starikov82}: 

\begin{equation}
\frac{\lambda_n}{\lambda_0} = \left[ \frac{1}{1 + \beta^2/2 + \beta\sqrt{(\beta/2)^2+1}} \right]^n
\label{eigendistribution}
\end{equation}
with $\beta=\sigma_{\mu}/\sigma_I$. Using the geometrical series: 
\begin{equation}
\sum_{n=0}^{\infty} q^n = \frac{1}{1-q}~~\textnormal{when}~~q<1
\end{equation}
is possible to obtain the $CF$:

\begin{equation}
CF = \frac{1}{1+1/[\beta^2/2 + \beta \sqrt{(\beta/2)^2+1}]}
\end{equation}

\begin{figure}
  \centering
  \includegraphics[width=0.90\textwidth]{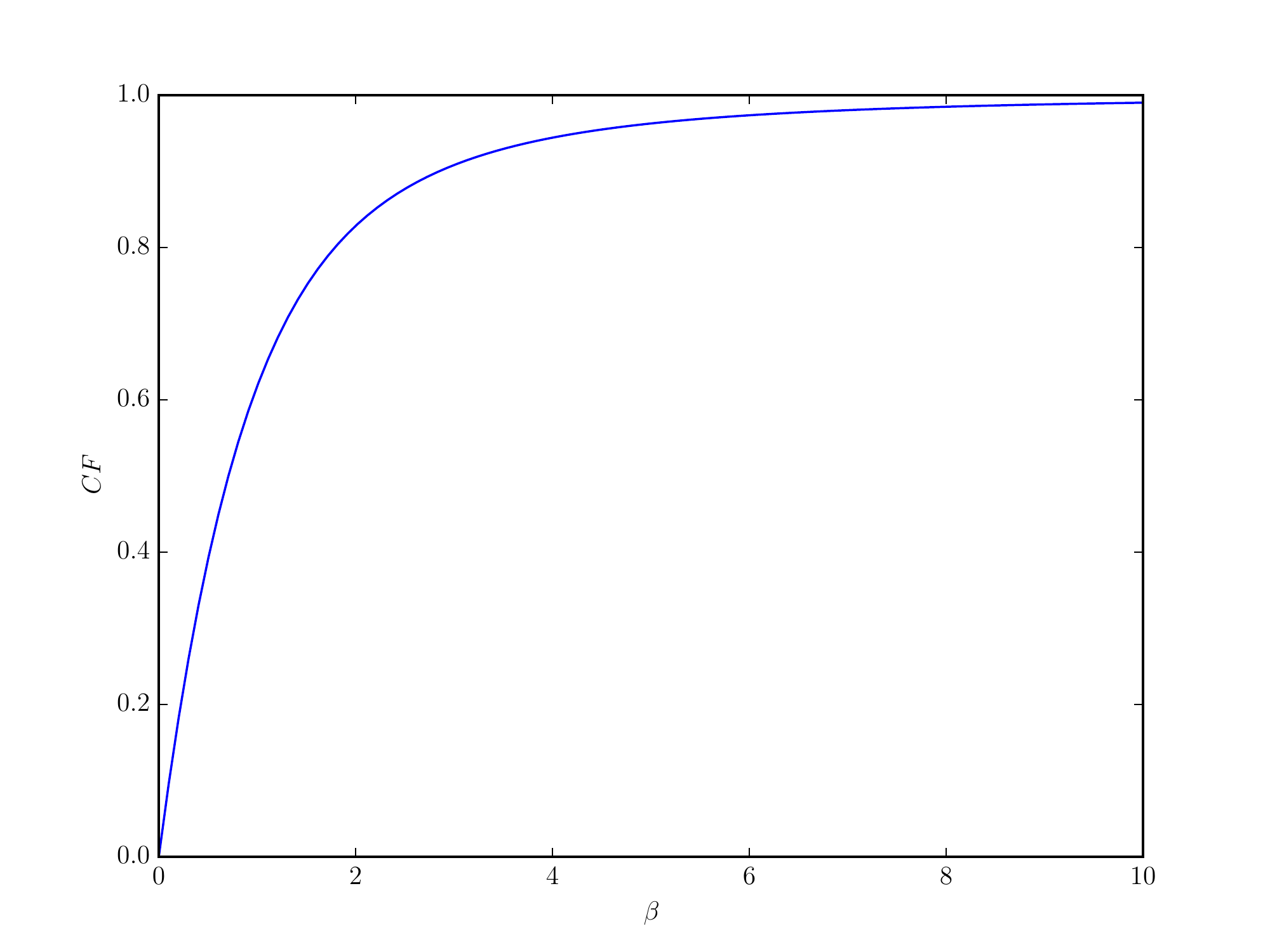} 
  \caption{Coherent fraction for the Gaussian Shell-model as a function of $\beta=\sigma_{\mu}/\sigma_I$}
  \label{fig:cf_gaussianshell}
\end{figure}

A plot of the $CF$ as a function of $\beta$ is in Fig.~\ref{fig:cf_gaussianshell}.

Some authors \cite{alsnielsen} arrive to the result similar to Eq.~\ref{sigmasigmaprime} in a completely different context by making assumptions starting from the Heisenberg principle. The Gaussian beam value of $(4 \pi)^{-1}$ in Eq.~\ref{sigmasigmaprime} is the minimum value that this space-beam-width product can have for any kind of optical beam, in analogy to the Gaussian minimum-uncertainty wave packets of Quantum Mechanics \cite{Siegman_1990}. Ref.~\cite{bazarov} makes a comparison of the radiation in quantum and classical mechanics using Wigner functions. I particularly think it is more rigorous to stay in a Classical Mechanics view and use Eq.~\ref{sigmasigmaprime} as a result of Classical Electrodynamics. It is however interesting to notice that using the Heisenberg principle one can see the factor $(4 \pi)^{-1}$ in Eq.~\ref{sigmasigmaprime} as the best possible condition.

\section{Undulator radiation}

Undulators are the most popular insertion devices used for producing synchrotron radiation. They produce radiation with non-smooth characteristics (spectral and spatial). The undulator magnets induce in the electrons mostly sinusoidal movement. The small deflection of the electron at each oscillation makes possible that the photons produced in a crest of the electron sinusoidal trajectory will interfere with the photons originated from the next oscillation crest, thus producing radiation whose spectrum contains peaks at photon energies proportional to the so-called resonance. The deflection parameter $K$ for an electron traveling in an oscillating magnetic field $B \cos(2 \pi z/\lambda_u)$ (with $z$ the spatial coordinate along the undulator, $B$ the maximum magnetic field, and $\lambda_u$ the undulator period) is:

\begin{equation}
K = \frac{e B \lambda_u}{2 \pi m c} \sim 93.3729 B[T] \lambda_u[m]
\label{K}
\end{equation}

And the wavelength of the resonance is:

\begin{equation}
\lambda = \frac{1+K^2/2+\gamma^2\theta^2}{2 \gamma^2} \lambda_u
\label{resonance}
\end{equation}
with $\gamma$ the electron energy in units of electron energy at rest, and $\theta$ is the observation angle ($\theta=0$ on-axis).

Undulator radiation can be calculated in the Classical Electrodynamics framework. The electric amplitude of the emission at a given point $r$ is obtained from an integral over the electron trajectory. The radiation can be calculated for any photon energy and results in a collimated cone of radiation at the photon energy of odd multipliers of the resonance energy. The spatial characteristics vary when shifting from the resonance and harmonic peaks. Most experiments exploit the fact of the high flux and good collimation of the beam at the resonance. The emission cone seen in the far field presents an intensity profile that does not change its shape when represented in angular coordinates, as a result of Fraunhofer propagation. The radiation emitted by a single electron (or filament beam) can be represented by a wavefront and is fully coherent. 

The width of the intensity profile of this radiation cone is the first element of discrepancy in literature. Krinsky, using simple arguments in Ref.~\cite{krinsky} affirms that the angular broadening of the radiation is defined as: 

\begin{equation}
\sigma_{r\prime} \cong \frac{1}{\gamma}\sqrt{ \frac{(1+K^2/2)}{2Nn}}=\sqrt{\frac{\lambda}{L} }
\label{kim_sigmaprime}
\end{equation}

The same definition is used by Kim in \cite{kim1986a,kim1986b} and also in the X-ray Data Booklet \cite{xraydatabooklet}. Notice that in the original texts the width is said to be 'half width', which in principle is different from 'sigma'.  In a Gaussian distribution the full width at half maximum (FWHM) is:

\begin{equation}
FWHM=2 \sqrt{2 \ln 2} \sigma \sim 2.355 \sigma
\end{equation}

In Kim's papers from 1989 \cite{kim1989}, the Gaussian approximation is obtained by comparing with the angular distribution of intensity (a {\it sinc} function) and obtain a rather smaller divergence:
\begin{equation}
\sigma_{r\prime} = \frac{1}{2\gamma}\sqrt{ \frac{(1+K^2/2)}{Nn}}=\sqrt{\frac{\lambda}{2L} }
\label{kimnew_sigmaprime}
\end{equation}

Elleaume \cite{elleaume} performs a Gaussian fit on the intensity versus emission angle at the resonance and obtains: 
\begin{equation}
\sigma_{r'} \cong 0.69 \sqrt{\frac{\lambda}{L}}
\label{elleaume_sigmaprime}
\end{equation}

\begin{figure}
  \centering
  \includegraphics[width=0.90\textwidth]{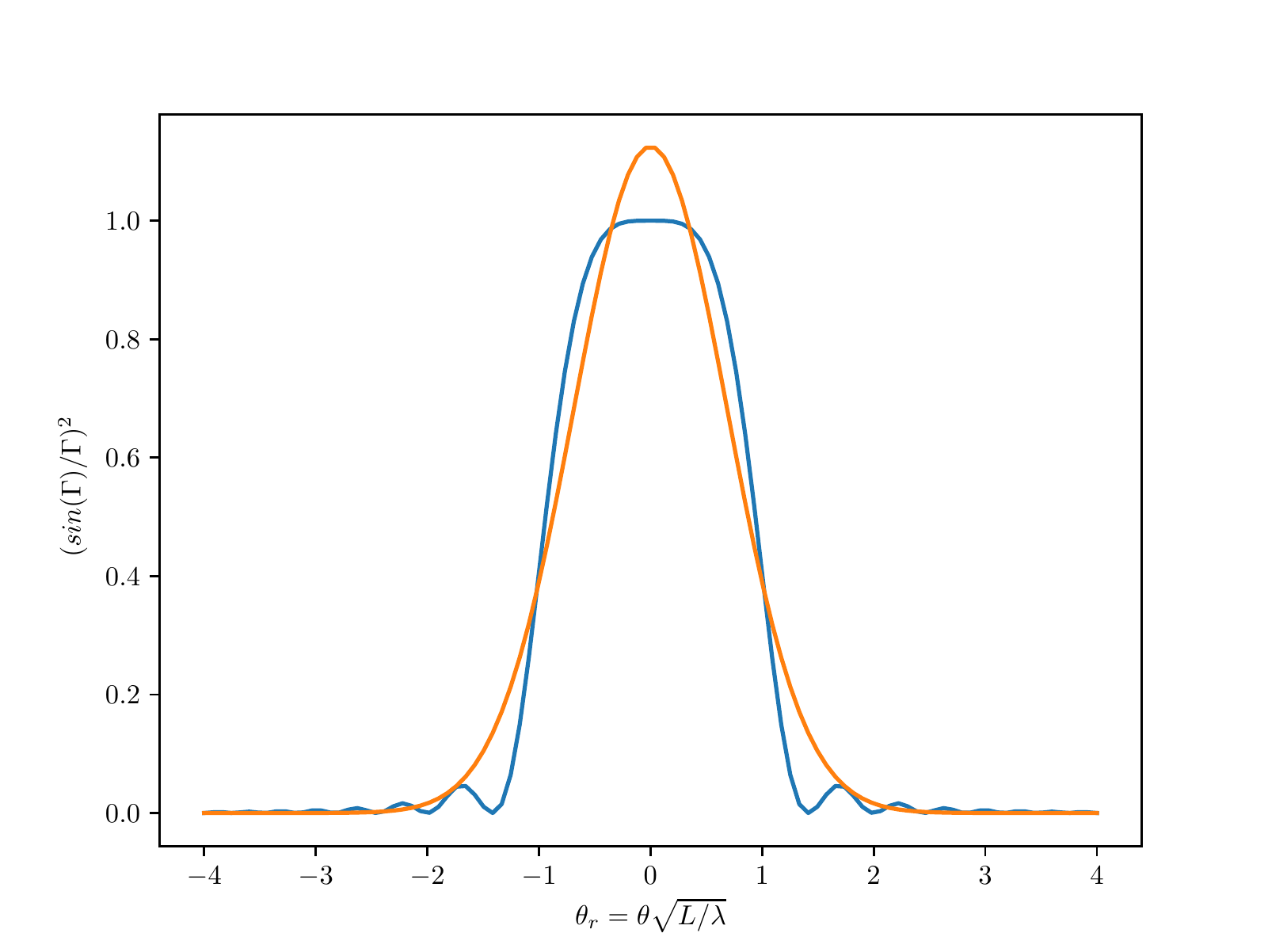} 
  \caption{Gaussian fit of the intensity versus reduced emission angle after Ref.~\cite{elleaume}}
  \label{fig:undulator_fit}
\end{figure}

A similar fit has been done here and the result is in Fig.~\ref{fig:undulator_fit}. The Gaussian obtained has a $\sigma=0.687$ as a function of $\theta_r=\theta\sqrt{L/\lambda}$ in very good agreement with \cite{elleaume}. Once again, one can remark that the intensity profile is far from being Gaussian. Elleaume approximates the numerically fitted value to 

\begin{equation}
\sigma_{r'} \cong 0.69 \sqrt{\frac{\lambda}{L}} \sim \sqrt{\frac{\lambda}{2L}} 
\label{elleaume_sigmaprimeapprox}
\end{equation}

Therefore, Kim's (new) and Elleaume's (approximated) results of emission divergence are smaller than Krinsky and Kim's (old) result by roughly 17\%.

In order to calculate the spatial size of the emission Kim supposed a Gaussian beam thus satisfying Eq.\ref{sigmasigmaprime} (Eq.~21 in \cite{kim1986b} or Eq.~6.37 in \cite{kim1989}). Therefore, it is implicitely assumed the validity of approximating undulator radiation by a Gaussian Shell-model, or, in other words, accepting that the emission at the resonance is Gaussian. 
From Eq.~28 in Ref.~\cite{kim1986b}: 
\begin{equation}
\sigma_r = \frac{1}{4 \pi} \sqrt{\lambda L} 
\label{kim_sigma}
\end{equation}
of, from Eq.~6.37 in Ref.~\cite{kim1989}: 
\begin{equation}
\sigma_r = \frac{1}{4 \pi} \sqrt{2 \lambda L} 
\label{kimnew_sigma}
\end{equation}

Elleaume in Ref.~\cite{elleaume} follows a different direction. He does not suppose that the observed approximately Gaussian divergence comes from the Fraunhofer propagation of a Gaussian beam as hypothesized by Kim, but obtains a numerical fit of the calculated radiation expressed as a function of the real space at the source position: 

\begin{equation}
\sigma_r \cong \frac{2.740}{4 \pi} \sqrt{\lambda L} 
\label{elleaume_sigma}
\end{equation}
which , again, can be approximated: 

\begin{equation}
\sigma_r  \sim \sqrt{\frac{\lambda L}{2 \pi^2}}
\label{elleaume_sigmaapprox}
\end{equation}

The fitted Gaussian for spatial (Eq.~\ref{elleaume_sigma}) and angular (Eq.~\ref{elleaume_sigmaprime}) representations of the undulator radiation are not related via Fourier transform (as supposed by Kim) but instead verify: 

\begin{equation}
\sigma_r  \sigma_{r'} \cong \frac{1.89 \lambda}{4 \pi} \sim \frac{\lambda}{2 \pi}
\label{elleaume_sigmasigmaprime}
\end{equation}
which can be interpreted as the phase space volume of undulator radiation is approximately twice the phase space volume of the first coherent mode of a Gaussian beam. Let us remark that this undulator emission is fully coherent as it is produced by a single electron, or a filament beam, or a zero emittance storage ring. 
%

In the following, we consider two approaches after Kim and Elleaume. Both use the same angular width (Eqs. \ref{elleaume_sigmaprimeapprox} and \ref{kimnew_sigmaprime}). Kim obtains the spatial width from ideas compatible with the  Gaussian Shell-model (Eq.~\ref{kimnew_sigma}) but Elleaume used a direct fit value (Eq.~\ref{elleaume_sigmaapprox}). Thus the phase space volume of undulator emission is twice for Elleaume (Eq.~\ref{elleaume_sigmasigmaprime}) than for Kim (Eq.~\ref{kim_sigmaprime}). 

In literature, it can be found papers that follow Elleaume's (e.g.,\cite{borland2012,hettel2014}) and Kim's model (e.g., \cite{huang2013}). It is worth mentioning, citing \cite{elleaume}, that {\it these are approximations and should not be considered as fundamental results.} 

\section{Coherent fraction}

The coherent fraction is usually thought of (although never defined) as the volume of the phase space for the photon beam emitted by a singe electron (or filament beam) $\sigma_r \sigma_{r\prime}$ over the phase space volume of the photon beam emitted by the emittance ring with finite emittance. The value of the phase space volume of the non-zero emittance ring is usually calculated by adding in quadrature the electron contribution with the single photon emission:

\begin{equation}
\Sigma_{x,y}  = \sqrt{ \sigma_{x,y}^2 + \sigma_{r}^2 }
\end{equation}
\begin{equation}
\Sigma_{x',y'}  = \sqrt{ \sigma_{x',y'}^2 + \sigma_{r'}^2 }
\end{equation}

Again, to justify the addition in quadrature, it has been implicitly supposed that both components have Gaussian distribution, which is probably a good approximation for the electron beam, but not necessarily for the photon beam. 

As the phase space of the photon beam is physically limited by the single electron emission, it makes no sense to build zero-emittance storage rings (if it were possible). A storage ring that would satisfy  $\Sigma_{x,y}  \gtrsim  \sigma_{r}$ and $\Sigma_{x',y'}  \gtrsim  \sigma_{r'}$ (or in other words $\sigma_{x,y} \ll \sigma_{r}$ and $\sigma_{x',y'} \ll \sigma_{r'}$) for all photon wavelengths of interest would be {\it optimum} and the term {\it diffraction limited storage ring} could be appropriated. The fact is that this term is often used for planned new storage rings where  $\sigma_{x,y} \gtrsim \sigma_{r}$ and $\sigma_{x',y'} \gtrsim \sigma_{r'}$. The author considers the term {\it low emittance} or {\it ultra-low emittance storage rings} is more adequate for these rings than diffraction limited storage rings. 

Considering now the coherent fraction as the product of the horizontal and vertical phase space volume ratios, the coherent fraction can be written as:

\begin{equation}
CF = \frac{(\sigma_r \sigma_{r\prime})^2 }{ \sqrt{\sigma_x^2+ \sigma_r^2}  \sqrt{\sigma_{x'}^2+ \sigma_{r\prime}^2} \sqrt{ \sigma_y^2+ \sigma_r^2}  \sqrt{\sigma_{y'}^2+ \sigma_{r'}^2}}
\end{equation}
Replacing values by those given by Kim (Eqs.~\ref{kimnew_sigma} and \ref{kimnew_sigmaprime}): 
\begin{equation}
CF^K = \frac{
\left( \frac{\lambda}{4 \pi} \right)^2}
{\sqrt{ 
\left(\sigma_x^2 + \frac{\lambda L}{8 \pi^2} \right) 
\left(\sigma_y^2 + \frac{\lambda L}{8 \pi^2} \right)
\left(\sigma_{x'}^2 + \frac{\lambda}{2 L} \right)
\left(\sigma_{x'}^2 + \frac{\lambda}{2 L} \right)
}}
\label{cf_kim}
\end{equation}

And Elleaume's approximated values (Eqs.~\ref{elleaume_sigmaapprox} and \ref{elleaume_sigmaprimeapprox}): 
\begin{equation}
CF^{EA} = \frac{
\left( \frac{\lambda}{2 \pi} \right)^2}
{\sqrt{ 
\left(\sigma_x^2 + \frac{\lambda L}{2 \pi^2} \right) 
\left(\sigma_y^2 + \frac{\lambda L}{2 \pi^2} \right)
\left(\sigma_{x'}^2 + \frac{\lambda}{2 L} \right)
\left(\sigma_{x'}^2 + \frac{\lambda}{2 L} \right)
}}
\label{cf_elleaume_approx}
\end{equation}
For comparison purposes, it is also interesting to define $CF^E$, or the coherent fraction calculated using Elleaume's numerical values of the size and divergence (Eqs.~\ref{elleaume_sigma} and \ref{elleaume_sigmaprime}).  

Some values for existing ESRF and new ESRF-EBS under construction storage rings are in Table~\ref{table:sources}.

\begin{table}
\caption{Electron parameters at center of straight section for different storage rings.}
\begin{tabular}{lcccc}      
 Storage ring    & $\sigma_x$ [$\mu m$]       & $\sigma_y$ [$\mu m$]      & $\sigma_{x'}$  [$\mu$ rad] & $\sigma_{y'}$  [$\mu$ rad]    \\
\hline
 ESRF (High$\beta$)      & 387.8 & 3.5 & 10.3  & 1.2   \\
 ESRF (Low$\beta$)       & 37.4  & 3.5 & 106.9 & 1.2   \\
 ESRF-EBS (S28A lattice) & 27.2   & 3.4  & 5.2   & 1.4    \\
 ESRF-EBS (S28D lattice) & 30.18  & 3.64 & 4.37  & 1.37   \\
\end{tabular}
\label{table:sources}
\end{table}

\begin{figure}
  \centering
  \includegraphics[width=0.90\textwidth]{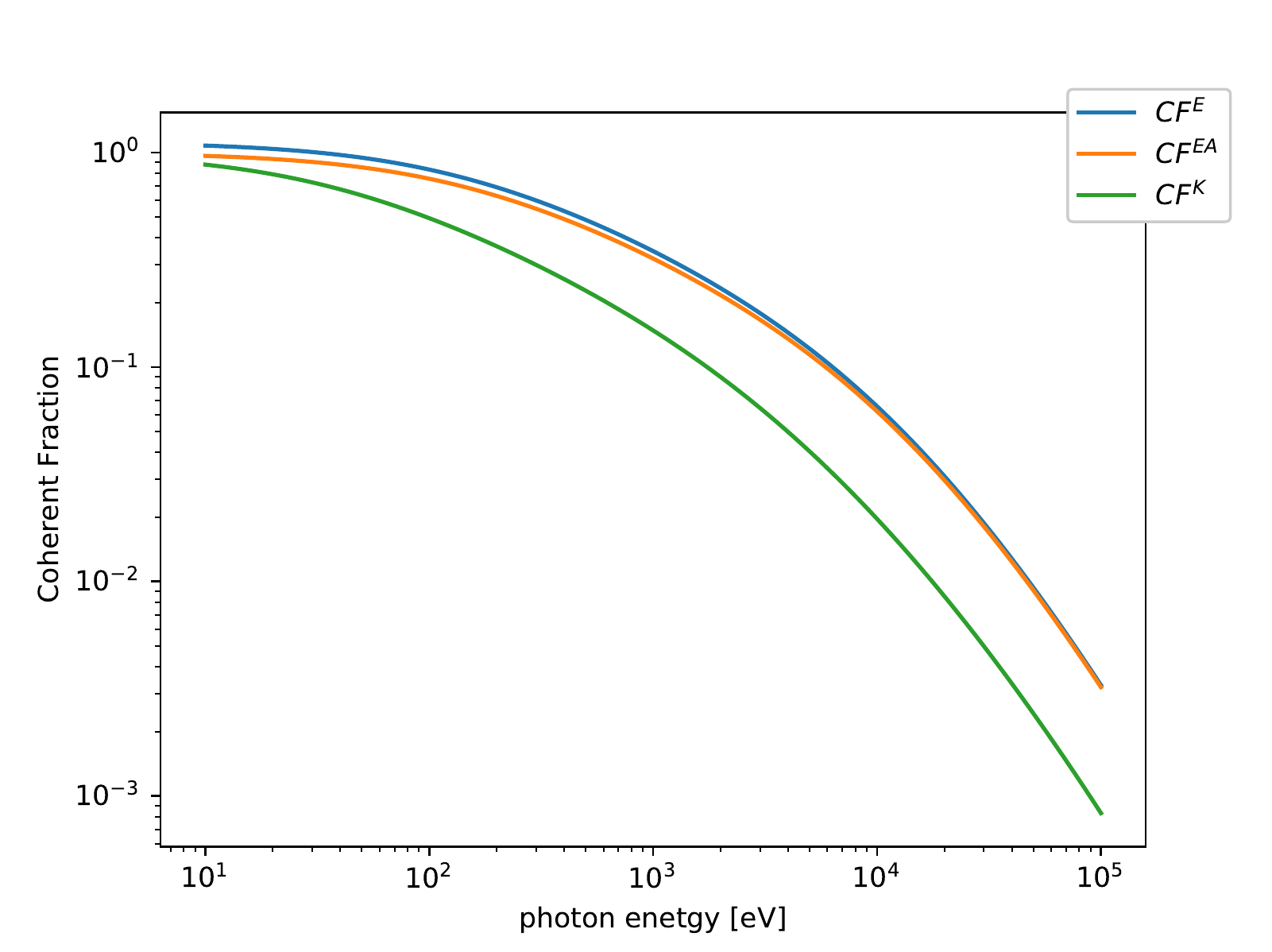} 
  \caption{Coherent fraction for ESRF-EBS (S28D) as calculated from Eq.~\ref{cf_kim} ($CF^K$),\ref{cf_elleaume_approx} ($CF^{EA}$) and coherent fraction $CF^E$ using  Elleaume's fit from Eqs.~\ref{elleaume_sigma} and \ref{elleaume_sigmaprime}.}
  \label{fig:CF}
\end{figure}

Fig.~\ref{fig:CF} shows a plot of the coherent fraction calculated for the ESRF-EBS using the Eqs.~\ref{cf_kim},\ref{cf_elleaume_approx} and
the Elleaume's method with using the numeric fits (Eqs.~19 and 21).

\section{Coherent fraction from coherent modes}

As stated in section 3, coherent fraction can be defined as the occupation of the first coherence mode. In order to calculate that, one should make the coherence mode decomposition of the undulator radiation. The full decomposition cannot be done analytically without going into severe approximations like the Gaussian Shell-model, which is not appropriate for low emittance storage rings. A methodology to compute numerically such decomposition has been recently proposed and successfully implemented \cite{glass2017}. From the numerical decomposition one obtains directly the coherent fraction from the first eigenvalue. We used COMSYL \cite{comsyl} for performing simulations of the 4 m long U18 undulator places at the ESRF storage ring (either in a Low-$\beta$ or a High-$\beta$ section) \cite{glass2017}. Results are shown in Table~\ref{table:CFmark} and compared with the approximated values from Eqs.~\ref{elleaume_sigma} and \ref{elleaume_sigmaprime}. Another example (see Table~\ref{table:CFid16}) compares the coherent fraction from a U18 L=2~m undulator placed at the EBS and ESRF High$\beta$ lattices. 

\begin{table}
\caption{Coherent fraction values (in $\%$) calculated as occupation of the first coherent mode using the COMSYL software ($CF^C$), and compared with approximated expressions $CF^E$, $CF^{EA}$ and $CF^K$ (see text) for an undulator U18 L=2m at 8 keV.}
\begin{tabular}{lcccc}      
 Storage ring    & $CF^C$       & $CF^{E}$      & $CF^{EA}$ & $CF^K$    \\
\hline
 ESRF (EBS)              & 7.02  & 8.67 & 8.19  & 2.74   \\
 ESRF (High$\beta$)      & 0.34  & 0.41 & 0.39  & 0.13   \\
 ESRF (Low$\beta$)       & 0.38  & 0.47 & 0.45  & 0.15   \\
\end{tabular}
\label{table:CFmark}
\end{table}

\begin{table}
\caption{Coherent fraction values (in $\%$) calculated as occupation of the first coherent mode using the COMSYL software ($CF^C$), and compared with approximated expressions $CF^E$, $CF^{EA}$ and $CF^K$ (see text) for an undulator U18 L=1.4m at 17.225 keV.}
\begin{tabular}{lcccc}      

 Storage ring    & $CF^C$       & $CF^{E}$      & $CF^{EA}$ & $CF^K$    \\
\hline
 ESRF (EBS)              & 2.8  & 3.22 & 3.08  & 0.88  \\
 ESRF (High$\beta$)      & 0.13 & 0.14 & 0.14  & 0.04   \\
\end{tabular}
\label{table:CFid16}
\end{table}

From these results we can observe that the exact numerical calculation $CF^C$ of the coherent fraction is better approximated by $CF^{EA}$ (Eq.~\ref{cf_elleaume_approx}), so using for $\sigma_r$ and $\sigma_{r'}$ Elleaume's values (Eqs.~\ref{elleaume_sigmaapprox} and \ref{elleaume_sigmaprimeapprox}, respectively). Therefore, the author advises to use these formulas for quick estimation of the coherent fraction.



\ack{Acknowledgements}
A large part of this work comes from Mark Glass' Ph. D. thesis work, who is also acknowledged for helpful discussions and corrections of this manuscript. This paper was motivated from discussions with several colleagues, such as Riccardo Bartolini, Eric Wallen, Rafael Celestre, Luca Rebuffi and Ruben Reininger.










\end{document}